\documentclass[reprint]{revtex4-1} 

\usepackage{amsmath}  
\usepackage{amsfonts} 
\usepackage{graphicx} 
\usepackage{hyperref}

\begin{document}

\title{Explosion analysis from images: Trinity and Beirut}
\author{Jorge S. D\'\i az}
\affiliation{Physics Department, Indiana University, Bloomington, IN 47405, U.S.A.}

\date{August 04,  2021}

\begin{abstract}
Images of an explosion can be used to study some of its physical properties.
After reviewing the key aspects of the method originally developed to study the first nuclear detonation and
analyzing the Trinity blast data,
the method is applied to the Beirut explosion of August 2020 by using images from videos posted online.
The applicability of the method is discussed and the process of selection, extraction, and analysis of the data is presented.
The estimate for the energy yield of the Beirut explosion is found to be $2.3^{+1.1}_{-1.1}$ TJ or $0.6^{+0.3}_{-0.3}$ kt of TNT equivalent.
The result is consistent with others recently appeared in the literature using different methods.
Notice that this article includes content that some readers may find distressing.
\end{abstract}
\maketitle 

\section{Introduction}

The mathematical description of an explosion size as a function of all the relevant physical quantities has become a classic in undergraduate classrooms over the years as a way of introducing dimensional analysis.
The story goes that, 
only using dimensional analysis, 
physicist Geoffrey I. Taylor was able to determine the yield of the first nuclear explosion (Trinity test).
This over-simplified account is usually accompanied by many other inaccuracies that have been preserved and exaggerated leading to a more dramatic narrative.
Some versions incorrectly portrait Taylor as an independent researcher that was not involved in the Manhattan Project; 
whereas others even suggest that Taylor revealed to the public a secret number from declassified information.
These misleading versions only add an entertainment aspect to an already scientifically interesting story that can be used in classrooms to show our students that the techniques they learn in their first courses can have real-world applications and even crucial consequences.
Taylor was not an independent researcher, 
his first report was the result of a request from the UK Ministry of Home Security in 1941 that shared highly classified information about the potential development of a fission-powered weapon, as narrated by himself \cite{Taylor1}.
During the Manhattan Project he was, 
together with Niels Bohr, 
one of the highly  distinguished  consultants that were made available under British auspices as part of the British mission to Los Alamos \cite{Fakley} and one of the selected group of scientists invited to the Trinity test \cite{Bainbridge}. 
Furthermore,
his now popular work determining the yield of the Trinity explosion from declassified images was published in 1950 \cite{Taylor1, Taylor2},
only after his two technical reports on the blast formation were declassified by the U.S. Atomic Energy Commission.
At this time the yield of the Trinity test as well as the two bombs dropped over Japan were already of public knowledge.

Nonetheless,
most of the inaccuracies appear in the evaluation of a dimensionless factor that cannot be accounted by dimensional analysis.
Many of the myths behind this story have been examined in detail by Deakin \cite{Deakin}, including comparisons between the work of Taylor \cite{Taylor1, Taylor2} with lesser known developments in parallel that took place in the U.S. by John von Neumann \cite{JvNeumann} and in the Soviet Union by Leonid Sedov \cite{Sedov}.

This article is organized as follows.
The general presentation of the evolution of an explosion is discussed in Sec. \ref{Sec:Description of the blast size}; 
the method used by Taylor is presented in Sec. \ref{Sec:Taylor}, 
and the application of the method to the data from the Trinity explosion is shown in Sec. \ref{Sec:Trinity}. 
The Beirut explosion and the application of the method to this event are described in Sec. \ref{Beirut},
including the selection of images, the steps followed for preprocessing, and the estimate of the energy yield.
Comparison and agreement with other methods recently published are also discussed.

\section{Description of the blast size}
\label{Sec:Description of the blast size}

The popular presentation of Taylor's work follows the description of a spherically symmetric explosion characterized by its radius $R$ in terms of the energy of the explosion $E_0$ released in a fluid of undisturbed density $\rho_0$, at a time $t$ since the detonation.
The assumption is that these quantities are related by power laws
\begin{equation}
R = S(\gamma)\,E_0^a \rho_0^b t^c,
\label{R(abc)}
\end{equation}
where $a$, $b$, and $c$ are dimensionless constants.
The dimensionless function $S(\gamma)$ has to be determined from the thermodynamical evolution of the explosion and it depends on the adiabatic index of the medium $\gamma$.
All popular accounts of Taylor's story completely ignore this observation and simply assume that the dimensionless quantity is a constant.
Most versions then assume that this constant is approximately 1; 
others go further and describe the tale of Taylor experimenting with small-scale explosions to determine the {\it constant}. 
This tale probably arose from an addendum that Taylor included at the end of his first paper at the time of declassification (1949), 
in which he briefly compares his theoretical description with newly available data of pressure measurements from the conventional explosion of RDX and TNT \cite{Taylor1}.

At this point dimensional analysis is introduced. 
The radius has units of length $[R] = L$, whereas the dimensions of the quantities on right-hand side are $[S(\gamma)]=1$, $[E_0]=ML^2T^{-2}$, $[\rho_0]=ML^{-3}$, and $[t]=T$.
The consistency of (\ref{R(abc)}) implies a system of linear equations relating the three exponents as follows:
\begin{eqnarray}
0 &=& a + b, \nonumber\\
1 &=& 2a - 3b, \\
0 &=& -2a + c,\nonumber
\end{eqnarray}
whose solution $a=-b=1/5, c=2/5$ allows writing (\ref{R(abc)}) in the form
\begin{equation}
R = S(\gamma)\left(\frac{E_0 t^2}{\rho_0}\right)^{1/5}.
\label{R}
\end{equation}
This is the first equation in Taylor's first paper \cite{Taylor1}.
It should be emphasized that more than simple application of dimensional analysis, 
the relationship in (\ref{R}) was formally obtained via a scale-invariance argument that reduced a system of partial differential equations into ordinary differential equations \cite{Bethe}. 
The point-source description of the blast by Seldov, Taylor, von Neumann assumes an instantaneous release of high energy from an infinitesimally localized source, assumptions that are perfectly valid for a nuclear explosion.
Another critical misconception of Taylor's work is solving the last equation for the energy in the form
\begin{equation}
E_0 = \frac{\rho_0 R^5}{S(\gamma)^5 t^2},
\label{E}
\end{equation}
because this apparently shows that a single measurement of the fireball size $R$ at time $t$ after the detonation suffice to determine the energy.
This assumes that the functional relationship between all the quantities involved is correct; unfortunately, a single measurement cannot provide any information about the validity of this assumption.
Several measurements in the form of pairs $(t,R)$ are needed to first verify the validity of  (\ref{R}),
and then the energy can be determined.
This is exactly what Taylor did in his second paper \cite{Taylor2}.

\section{Taylor's method}
\label{Sec:Taylor}

In 1941, G. I. Taylor developed a theoretical description of the formation of a blast by a hypothetical nuclear explosion \cite{Taylor1}.
His report remained classified until 1949.
The day after the second anniversary of the Trinity test, 
the U.S. Atomic Energy Commission declassified a technical report including 25 images of the Trinity explosion indicating timestamps and a length scale \cite{Mack}.
This led Taylor to write his now-famous second paper, in which he makes use of the 25 available pairs $(t,R)$ to assess his theoretical formulation.
In early laboratory experiences students learn that the analysis of power-law and exponential relations can be done with ease by taking advantage of the properties of logarithms,
that allow converting exponents into simple multiplicative factors.
This is how students can analyze the terminal velocity of falling objects in Mechanics class and the time constant in RC circuits in Electromagnetism class:
the determination of exponential and other factors is reduced to simply estimate the slope and intercept of a straight line from their experimental data in log-log space.
Taylor followed this method to reduce the complexity of the relationship between $R$ and $t$ in  (\ref{R}),
which can be written as
\begin{equation}
\frac{5}{2} \log R = \log t + \frac{1}{2} \log{\left({\frac{S(\gamma)^5 E_0}{\rho_0}} \right)}.
\label{logR}
\end{equation}
This expression indicates that if instead of the pairs $(t,R)$ we use the pairs $(x,y)=(\log{t},5/2 \log{R})$, 
then the plot will be a straight line of the form $y(x) = mx + n$,
with $m=1$.
Here Taylor makes two clear predictions: 
\begin{enumerate}
\item the pairs $(\log{t},5/2 \log{R})$ will follow a straight line with slope 1;
\item the intercept $n$ can be used to determine the energy using
\begin{equation}
n = \frac{1}{2} \log{\left({\frac{S(\gamma)^5 E_0}{\rho_0}} \right)}.
\label{n(E)}
\end{equation}
\end{enumerate}
Notice that prediction 2 requires all pairs to be along the same line so that there is a unique value for the intercept $n$.
The high temperatures involved would lead to changes in the value of $\gamma$ due to  the increase of $C_V$ via absorption of energy in the form of molecular vibrations of the gases in the air as well as the absorption of intense radiation in the outer layers of the blast.
This means that the data could satisfy prediction 1 but not necessarily prediction 2 due the functional dependence of the factor $S(\gamma)$ on the fluctuating adiabatic index.
The validity of the two predictions implies from  (\ref{n(E)}) that the energy yield of the explosion can be written
\begin{equation}
E_0 = \frac{10^{2n} \,\rho_0}{S(\gamma)^5}.
\label{E(n)}
\end{equation}

\begin{figure}
\centering
\includegraphics[width=3.3in]{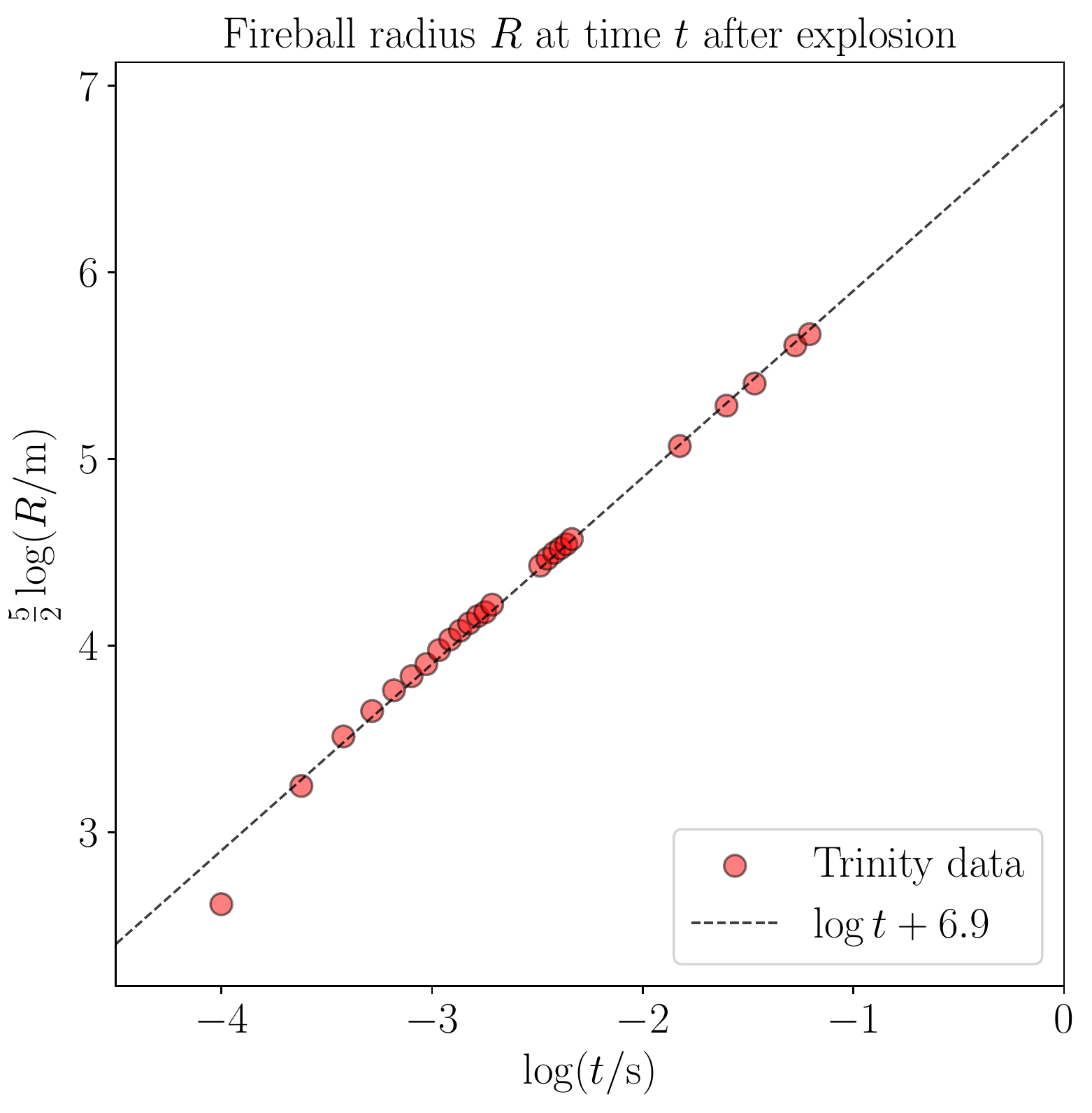}
\caption{Reproduction of Taylor's logarithmic plot showing that the data from the Trinity explosion follows the predictions of  (\ref{logR}). 
The dashed line corresponds to the linear fit.
Notice that Taylor used CGS units \cite{Taylor2}.}
\label{Fig:tR_plot_Trinity}
\end{figure}

\section{Trinity explosion}
\label{Sec:Trinity}

From the technical report declassified by the U.S. Atomic Energy Commission \cite{Mack} Taylor constructed a table with the $(t,R)$ and $(\log{t},5/2 \log{R})$ pairs \cite{Taylor2}.

The result is reproduced in Figure \ref{Fig:tR_plot_Trinity}, 
which shows a remarkable agreement of the data with the theoretical description in  (\ref{logR}).
A simple linear fit results in the parameters $m = 1.0\pm0.0$ and $n = 6.9\pm0.1$,
with all the data following a single line confirming the two predictions of Taylor's modeling.
This last observation suggests that all the effects leading to variations in the value of the adiabatic index interplay producing a constant value for the effective $\gamma$ \cite{Taylor2}.
Considering an atmospheric explosion and the diatomic nature of nitrogen and oxygen that compose most of the air, 
the adiabatic index is $\gamma=7/5$.
Taylor computed the numerical value of $S(\gamma)^{-5}$ for different situations.
For $\gamma=7/5\approx1.4$ he found $S(1.4)^{-5}=0.856$ (denoted $K$ in the second paper) \cite{Taylor2}.
Notice that this implies $S(1.4)=1.032$, which coincidentally is close to unity and makes those misleading versions of Taylor's story discussed in the previous section produce results in close agreement with the formal analysis.
Using the density of air $\rho_0=1.23$ kg/m$^3$ and the intercept from the data produces the estimate for the yield of the first nuclear explosion to be $E=66\pm31$ TJ.
Using the convention that a million kilograms of TNT (kt) corresponds to $4.18$ TJ \cite{nist},
the data in Figure (\ref{Fig:tR_plot_Trinity}) leads to a yield of $16\pm7$ kt.
This value is remarkably close to the reported estimated yield of 20 kt based on direct measurements of the pressure wave of the blast during the Trinity test.
Although the reported official value of the yield fluctuates within the range 15--20 kt,
the result from the 25 images confirms the outstanding validity of Taylor's modeling of the explosion during the first dozens of milliseconds.

\begin{figure}
\centering
\includegraphics[width=3.3in]{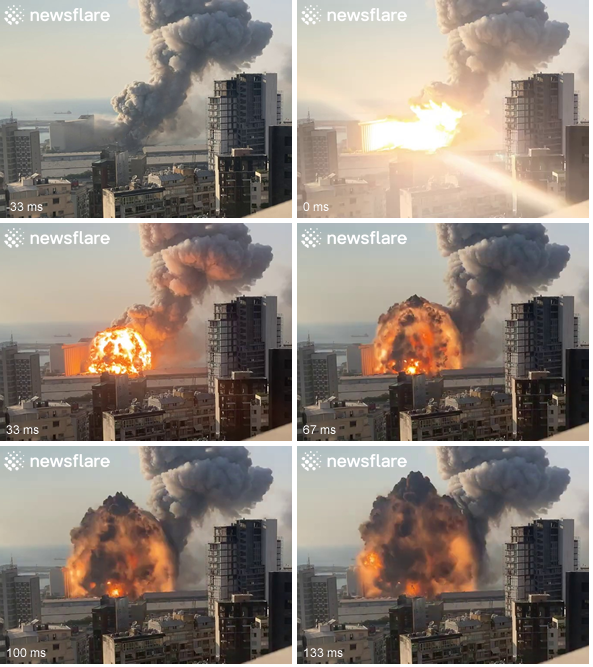}
\caption{First frames of the expanding fireball from Video 2. 
Footage courtesy of Newsflare Ltd.}
\label{Fig:frames}
\end{figure}

\section{Beirut explosion}
\label{Beirut}

On August 4, 2020 a devastating explosion occurred at the port of Lebanon's capital Beirut.
Preliminary reports indicate that a fire affected a warehouse where fireworks were stored together with ammonium nitrate.
The main explosion was preceded by an early deflagration of the firework products generating a thick gray column of smoke.
This led to many video recordings of the events that followed from multiple angles around the city.
Many videos show the main explosion as an expanding fireball surrounded by a short-lived Wilson cloud that engulfed the buildings near the port before disappearing and making visible an ascending red-brown mushroom cloud  characteristic of ammonium nitrate explosions,
followed by the corresponding blast wave that caused great damage.
Early reports suggest that the blast was caused by the detonation of 2750 metric tons of ammonium nitrate stored in one of the port's warehouses \cite{ANnature}. 

One feature that captured a fair amount of attention and that is clearly visible in all videos is the mentioned Wilson cloud that expanded and rapidly disappeared after the main blast.
If the air around the explosion has a high content of water-droplet aerosols then the pass of the pressure wave can become visible.
The sudden drop in pressure behind the pressure wave briefly prompts the relative humidity to supersaturation,
which dramatically enlarges the size of the droplets producing a visible cloud \cite{Waltz}.
Most media outlets incorrectly called this a `mushroom cloud' because of the similarity of historical footage of nuclear tests in the ocean.
The column of dark smoke that followed the explosion did produce an actual mushroom cloud;
nevertheless,
there was a confusion between the smoke mushroom cloud and the semi-spherical Wilson cloud.

Footage of the explosion circulated rapidly via social-media channels.
Given the availability of video footage, such as Figure \ref{Fig:frames}, the yield of the Beirut explosion could be estimated using Taylor's method.
Nonetheless, the reader is warned that the results in the previous section might not be applied to a chemical explosion because the assumptions of the point-source solution are not necessarily satisfied.
There is a window of values for the reduced blast size $Z=R/E_0^{1/3}$ for which the point-source solution is still applicable to chemical explosions and, as we will see, the early stages of the Beirut blast satisfy this condition making Taylor's method applicable.

\subsection{Methodology}
In order to apply Taylor's method to the Beirut explosion we must construct a dataset of pairs $(t, R)$, where the radius of the explosion $R$ at a time $t$ is precisely measured from the publicly available videos.

\subsubsection{Time measurements.}
A frame-by-frame analysis allows a clear identification of the first 200 milliseconds of the expanding fireball.
It is important to note that the fireball and the shock wave are different features of the explosion, in fact, the speed of the fireball quickly declines, whereas the blast wave continues rapidly expanding.
For this reason, we will restrict the analysis only to the early stages when fireball and blast wave can be approximated to have a similar size.
In the scenario of an early decoupling between fireball and blast wave, the results obtained in the following analysis can be interpreted as a lower bound on the energy released because we will use images of the clearly visible and slower fireball only.
Most commercial phones record video at a standard of 30 frames per second,
leading to the capture of around six frames of the early stages of the fireball.
The value of $t=0$ is identified as the frame in which the flash of the explosion is visible in each video 
and the inverse of the frame rate (1/FPS) serves as an assessment of the uncertainty of the time measurements.

\subsubsection{Fireball size measurements.}
\label{Sec.R}
The next step in the construction of the set of pairs $(t, R)$ is a method to find the size of fireball.
We could measure the opening angle occupied by the fireball in the field of view of the frame and then use the distance from the recording location to ground zero and apply standard trigonometric relations.
However, this procedure relies on the accuracy of online tools for measuring distances and could introduce unexpected uncertainties in the size measurements.
Another idea is proposed as follows:
the explosion took place at Warehouse 12 located adjacent to the 150-meters-long grain silos visible in all videos;
a simple proportionality relation between pixels and physical length can be established using the known dimensions the grain silos.
For each video,
a reliable conversion factor can be obtained by identifying the location from where the footage was recorded,
which allows determining the angle between the line of sight and the main axis of the rectangular grain silos building.
This angle is then used to find the projection of the building visible from the camera point of view in meters,
which can be related to the corresponding view in pixels.
The final step is measuring the size of the fireball in each frame in pixels and use the conversion factor to obtain the radius in meters.

\begin{figure}
\centering
\includegraphics[width=3.3in]{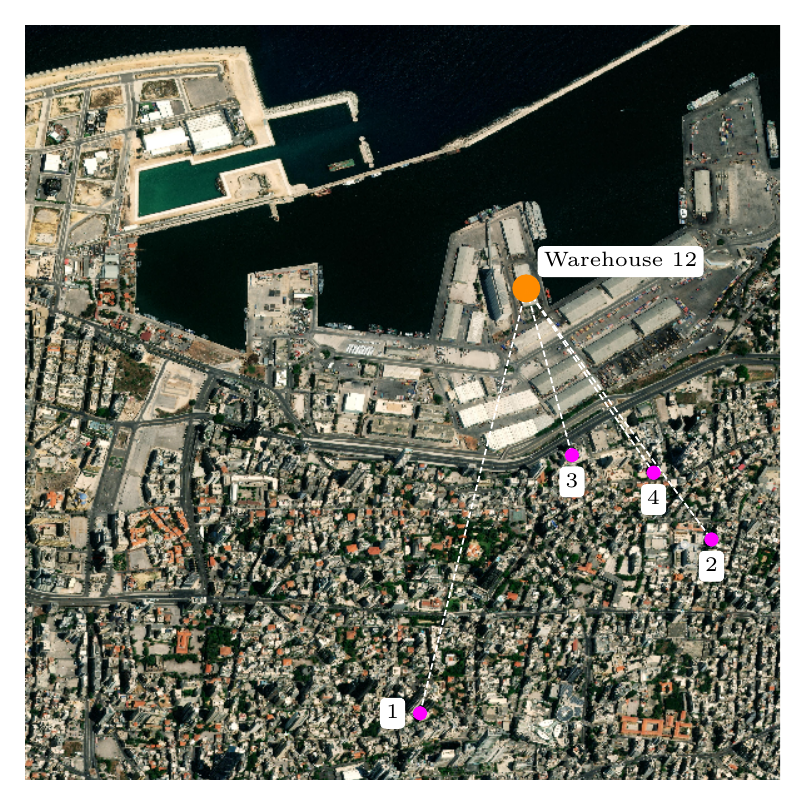}
\caption{Recording locations of the four selected videos.
Satellite image of Beirut on July 31, 2020 \cite{Maxar}.}
\label{Fig:AllViews4}
\end{figure}

\subsection{Video Selection}
Many videos and images were shared in social-media platforms before the explosions took place.
A complete reconstruction of the main events on the day of the blast \cite{ForensicArch} shows that the first image reporting the fire and smoke from Warehouse 12 was posted more than 10 minutes before the main explosion that occurred at 18:08 local time (15:08 UTC).
The detonation was captured from many angles and levels around the city and videos have been selected was based on the following criteria: 
\begin{enumerate}
\item enough of the surroundings are visible for a clear identification of the recording location;
\item there is a clear view of the grain silos to determine the conversion factor; 
\item at least four points of the evolution of the fireball are visible. 
\end{enumerate}
Out of the many videos publicly available,
only four met the criteria \cite{Darahagi1, Agoston, Alavi, Darahagi2}.
Their recording locations with respect to the site of the explosion are shown in Figure \ref{Fig:AllViews4}.

\subsection{Frame Extraction}
The frame extraction for each video was implemented using {\it OpenCV}, an open-source computer-vision library in Python that also allows determining the timestamps for each frame with millisecond precision \cite{OpenCV}.
Each video in MP4 format was processed using the OpenCV class for video capturing from video files and image sequences called \verb"VideoCapture".
The frame rate was obtained using the \verb"CAP_PROP_FPS" attribute of the video object, 
and the attribute \verb"CAP_PROP_POS_MSEC" was used to obtain the timestamps in milliseconds.
Finally, the method \verb"imwrite" was used to save each individual video frame.
The procedure was repeated for each video.

\subsection{Data Processing}
The determination of the size of the explosion for each frame was achieved by measuring the size of the fireball in pixels and then converting to meters using the corresponding factor for each video, as described in Sec. \ref{Sec.R}.
A fully automatic measurement of the fireball is challenging due to the thick dark smoke that appears in the first milliseconds covering significant fractions of the fireball.
The process was completed using a hybrid manual/semi-automatic method. 
The manual procedure consisted in examining each image using image-processing software by zooming into the region of interest and determining the center of the fireball and its approximate diameter.
This procedure accounts for most of the uncertainties.
By repeating measurements several times an approximate error of a few meters ($\leq8$ m.) was found.
Another source of uncertainty is the measurement of the angle between the line of sight and the main axis of the grain silos used to determine the conversion factor;
nevertheless,
this was found to be less than 2\% of the error from the fireball measurement.
The total uncertainty was rounded up to a moderate 10 m. in the analysis.
For the semi-automatic method,
the images were decomposed into their RGB channels for independent study.
In the first frames the information in R-channel dominates, while the B-channel is mostly opaque.
The brightness of the early fireball made the R-channel particularly useful.
The images were transformed to binary by setting individual pixels above a threshold to one and to zero otherwise.
This threshold was manually determined and the process was done using the methods
\verb"threshold" and \verb"THRESH_BINARY" in OpenCV.
The result was a collection of irregular patches that were smoothed using the \verb"erode" method.
Then the \verb"connectedComponents" method was used to find the patch of pixels corresponding to the fireball,
which was subsequently filled using the method \verb"floodFill".
This last step returns a single patch of pixels with the approximate shape of the fireball, which can be used to determine the center of the fireball using the method \verb"ndimage.measurements.center_of_mass" in SciPy \cite{SciPy}.
The described method confirmed the validity of the manual measurements of the fireball center and radius within the uncertainty $\delta R=$10 m. mentioned above.
A total of 26 pairs $(t,R)$ were obtained and are presented in Table \ref{table:data}.
Images in the form of arrays of data made use of NumPy \cite{NumPy} and the data was handled using pandas \cite{pandas}.

\begin{table}
\caption{Data extracted from the four selected videos.}
\footnotesize
\begin{tabular}{c c c c c c}
\hline
$R$ (px)	&	$R$ (m)	&	$t$ (s)	&	Video	&	$5/2 \log R$	&	$\log t$	\\ \hline
75	&	67.39	&	0.017	&	3	&	4.57	&	-1.77	\\
113	&	75.79	&	0.033	&	4	&	4.70	&	-1.48	\\
80	&	77.83	&	0.033	&	2	&	4.73	&	-1.48	\\
99	&	88.51	&	0.035	&	3	&	4.87	&	-1.46	\\
111	&	90.36	&	0.034	&	1	&	4.89	&	-1.47	\\
112	&	100.64	&	0.052	&	3	&	5.01	&	-1.28	\\
118	&	114.31	&	0.067	&	2	&	5.15	&	-1.17	\\
143	&	116.41	&	0.067	&	1	&	5.16	&	-1.17	\\
176	&	118.04	&	0.066	&	4	&	5.18	&	-1.18	\\
132	&	118.16	&	0.070	&	3	&	5.18	&	-1.15	\\
136	&	122.20	&	0.087	&	3	&	5.22	&	-1.06	\\
147	&	132.09	&	0.105	&	3	&	5.30	&	-0.98	\\
163	&	132.69	&	0.100	&	1	&	5.31	&	-1.00	\\
202	&	135.14	&	0.100	&	4	&	5.33	&	-1.00	\\
140	&	136.20	&	0.100	&	2	&	5.34	&	-1.00	\\
158	&	141.52	&	0.122	&	3	&	5.38	&	-0.91	\\
150	&	145.93	&	0.133	&	2	&	5.41	&	-0.88	\\
166	&	149.16	&	0.139	&	3	&	5.43	&	-0.86	\\
223	&	149.22	&	0.133	&	4	&	5.43	&	-0.88	\\
184	&	149.78	&	0.134	&	1	&	5.44	&	-0.87	\\
171	&	153.65	&	0.157	&	3	&	5.47	&	-0.80	\\
175	&	157.25	&	0.174	&	3	&	5.49	&	-0.76	\\
168	&	162.95	&	0.167	&	2	&	5.53	&	-0.78	\\
202	&	164.03	&	0.167	&	1	&	5.54	&	-0.78	\\
180	&	175.11	&	0.200	&	2	&	5.61	&	-0.70	\\
185	&	179.97	&	0.233	&	2	&	5.64	&	-0.63	\\
\hline
\end{tabular}
\label{table:data}
\end{table}

\begin{figure}
\centering
\includegraphics[width=3.3in]{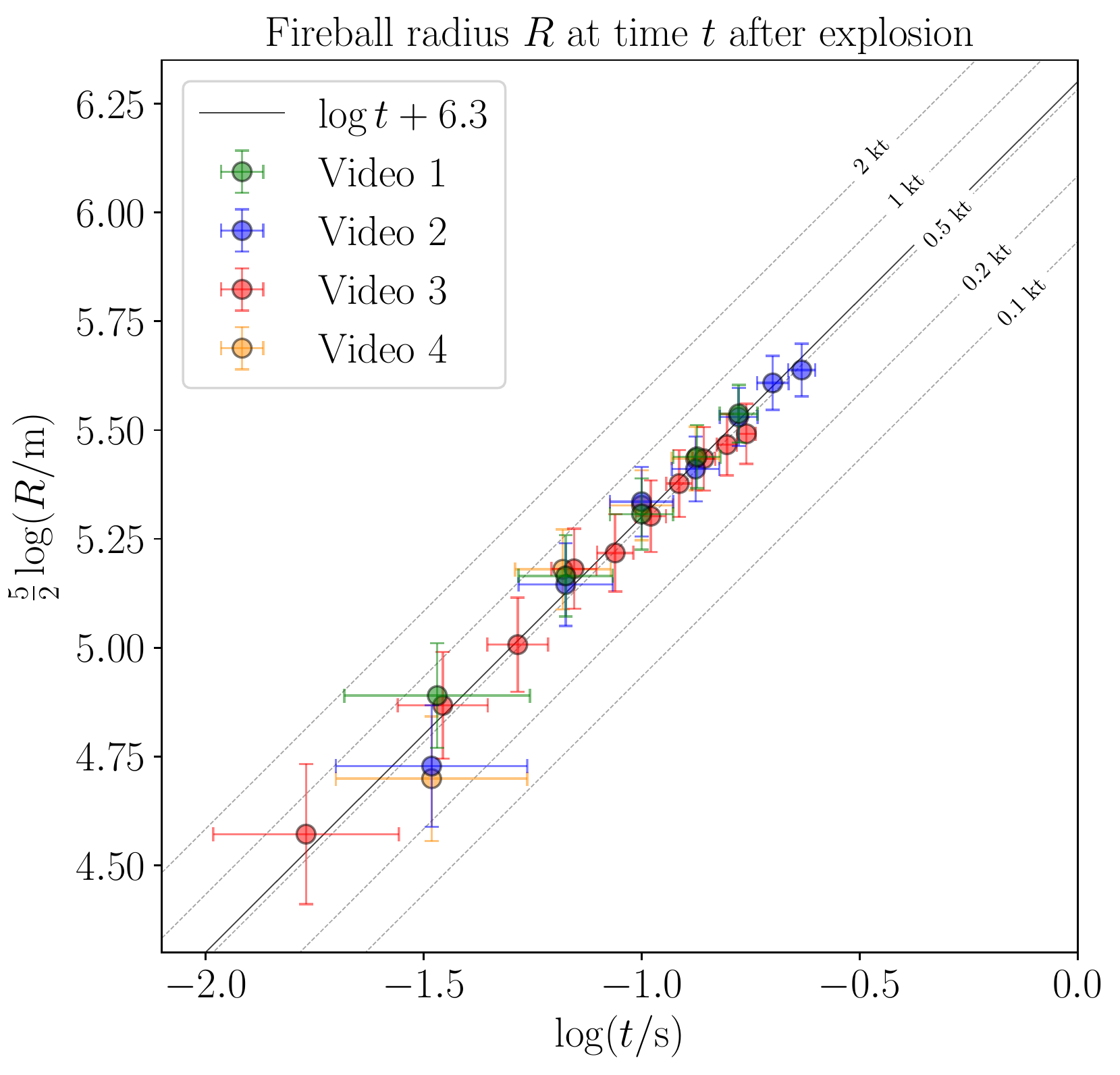}
\caption{Logarithmic plot showing the data from the four selected videos following the predictions of (\ref{logR}). 
The solid line corresponds to the median value of the distributions of fit parameters found in Sec. \ref{Results} and the dashed lines show the lines of 0.1, 0.2, 0.5, 1, and 2 kt of TNT equivalent produced by a hemispherical explosion for reference.
The labels indicate the recording locations in Figure \ref{Fig:AllViews4}.}
\label{Fig:tR_plot_Beirut}
\end{figure}

\subsection{Analysis and Results}
\label{Results}

\begin{figure}
\centering
\includegraphics[width=3.3in]{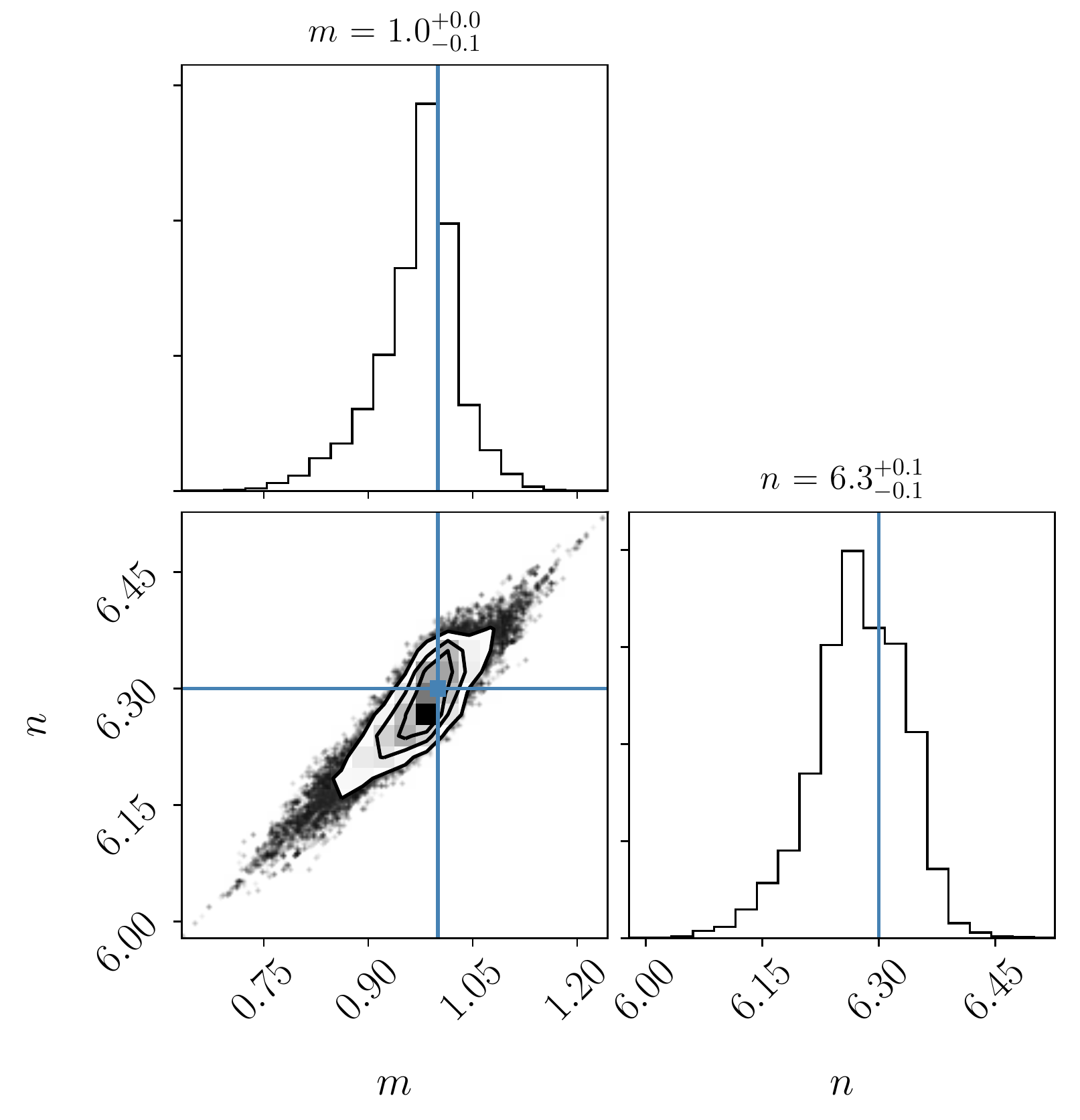}
\caption{One and two dimensional projections of the posterior probability distributions of the model parameters.
The lines indicate the median value of the distribution for each parameter.}
\label{Fig:MCMC}
\end{figure}

Following Taylor's method presented in Sec. \ref{Sec:Taylor},
the data in Table \ref{table:data} from the four selected videos is represented in Figure \ref{Fig:tR_plot_Beirut},
where the error bars show the uncertainty properly propagated as $\delta(\frac{5}{2}\log R) = \frac{5}{2}\frac{\delta R}{R\ln 10}$ and $\delta(\log t) = \frac{\delta t}{t\ln 10}$.
Remarkably,
all the data is consistent with a single straight line of unity slope,
as described by  (\ref{logR}).
The formulation presented in Sec. \ref{Sec:Taylor} assumes the energy to be released instantaneously from a single point,
whereas the ammonium nitrate in Warehouse 12 of Beirut's port was far from a point source;
moreover,
the chemical nature of the energy release makes it slower than the case of a nuclear detonation.
Nevertheless,
Figure \ref{Fig:tR_plot_Beirut} shows that for the first 200 milliseconds (\ref{logR}) describes the evolution of the fireball remarkably well.
A Bayesian approach for fitting Taylor's model to the data can be followed by using \verb"emcee",
a Python implementation of the affine-invariant ensemble sampler for Markov Chain Monte Carlo (MCMC) \cite{emcee, emcee2}.
The resulting projections of the posterior probability distributions of the model parameters are shown in Figure \ref{Fig:MCMC} \cite{corner}.

The projections of the posterior probability distributions of the slope $m$ and intercept $n$ show an excellent agreement with the unitary slope predicted by Taylor's model within the uncertainty of the data.
The slope is found to have the value $m = 1.0^{+0.0}_{-0.1}$,
whereas the intercept is given by $n = 6.3^{+0.1}_{-0.1}$.
The uncertainties are based on the 16th and 84th percentiles of the samples in the marginalized distribution for each parameter.

The final step of this analysis is to use (\ref{E(n)}) to determine the energy from the intercept $n$ found above.
Nonetheless,
there is a significant observation to consider: Taylor described an explosion in free air, which is perfectly valid for the Trinity test, in which the first nuclear bomb was detonated at the top of a 30-meter steel tower allowing the early stages of the explosion expand spherically.
The Beirut explosion, on the contrary, occurred at ground level;
therefore,
it is better described as the explosion of a hemispherical charge.
In practice,
the yield of a spherical charge needed to produce a blast of a given size is larger than the equivalent yield from a hemispherical charge.
If the ground were a perfect reflector of the blast wave, then the hemispherical explosion would appear enhanced by a factor two with respect to a spherical explosion in free air;
nevertheless,
around 10\% of the energy is dissipated as ground shock and cratering so that the enhancement factor is closer to 1.8 \cite{factor18}.
This means that the yield calculated in (\ref{E(n)}) must be corrected by a factor 1/1.8 to account for the hemispherical nature of the Beirut explosion \cite{footnote}.
Using (\ref{E(n)}), we find the estimate for the energy of the Beirut explosion to be $E_0=2.3^{+1.1}_{-1.1}$ TJ,
corresponding to a yield of $0.6^{+0.3}_{-0.3}$ kt of TNT equivalent.

\subsection{Applicability to a chemical explosion}
At the beginning of Sec. \ref{Beirut} the reader was warned about the direct application of a method developed for nuclear explosions to a chemical explosion.
Significant differences between these two type of explosions can lead to different phenomena, in particular to how the blast wave develops, a key assumption of the present study.
The point-source solution developed by Taylor, Seldov, and von Neumann relies on clear assumptions that are not necessarily satisfied by a chemical explosion.
Taylor explored the limitations of his formulation when applied to high explosives and using experimental data found that when the reduced blast size $Z=R/E_0^{1/3}$ lies in range $(2.2$--$3.8)\times10^{-3}$ cm/erg$^{1/3}$ then the point-source solution is still applicable \cite{Taylor1}.
Converting the length to meters and the energy to terajoules this range becomes $47.3$--$81.7$ m/(TJ)$^{1/3}$.
Using the median explosive yield determined in the previous section we find that the Beirut explosion lies within the range of applicability for distances between 62--108 meters.
Table \ref{table:data} shows a significant fraction of the data reaching beyond this range;
nonetheless,
even the data points outside the region of applicability remarkably follow the unique straight line corresponding to the median explosive yield estimated in the previous section.
A similar observation has been reported by Aouad et al. \cite{Aouad}.

\subsection{Comparison with other yield estimates}

The earliest estimate of the yield of the Beirut explosion was communicated by the Federal Institute for Geosciences and Natural Resources in Germany.
Using data registered by infrasound sensors from the International Monitoring System (IMS) of the Comprehensive Nuclear-Test-Ban Treaty (CTBT) as well as seismic and hydroacoustic signals recorded by seismometers Pilger et al. estimated an average explosive yield of 0.8--1.1 kt TNT equivalent \cite{Pilger}.
The first study making use of social-media videos applied a semi-empirical relation between time of arrival of the blast wave and distance from ground zero in terms of the explosive yield.
By fitting this relationship Rigby et al. found the best estimate 0.50 kt TNT with an upper limit of 1.12 kt TNT \cite{Rigby}.

\begin{figure}
\centering
\includegraphics[width=3.3in]{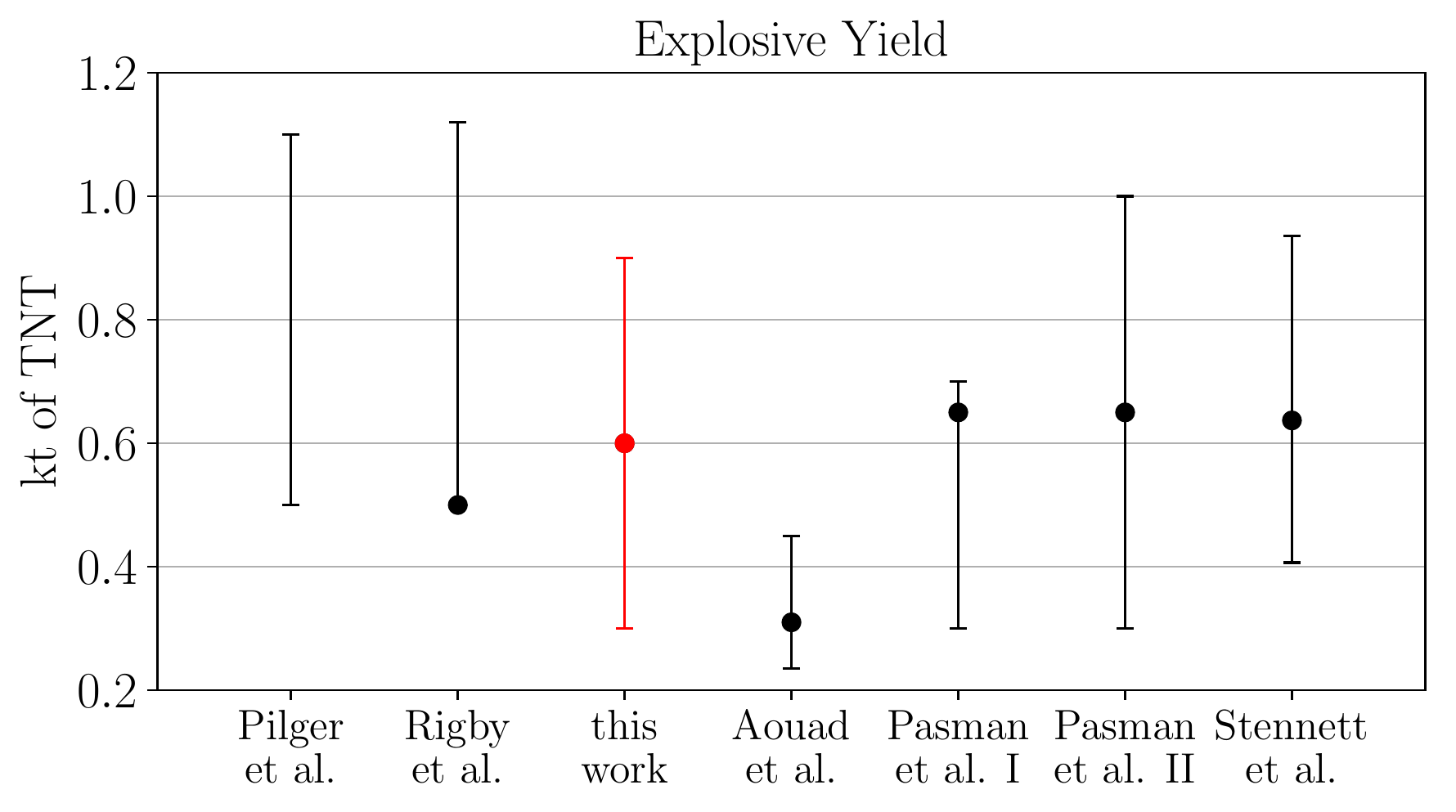}
\caption{Comparison of the estimated explosive yield by different studies recently published and the result reported in this work.
The circle represents the best estimate and the bars span between the lower and upper limits.}
\label{Fig:yield_comparison}
\end{figure}

During the review process of the present article other estimates have been reported that make use of social-media videos.
An independent report following the same procedures of the present work used six different videos to track the size of the fireball during the first 230 milliseconds of the blast.
Despite following a similar procedure the explosive yield obtained by Aouad et al. was $310^{+405}_{-235}$ tons of TNT \cite{Aouad}.
Similarly, Pasman et al. used videos and images to estimate the explosive yield using three different methods: a comparison with TNT curves, the crater dimensions, and the effects of the pressure wave based on visible damage.
Two ranges were reported 0.3--0.7 kt TNT and 0.3--1.0 kt TNT with the best estimate value at 0.65 kt TNT in both cases \cite{Pasman}.
Finally, using a publicly available blast calculator Stennett et al. estimated the yield of the explosion by relating the arrival time of the blast wave and the distance traveled leading to the range 0.407--0.936 kt TNT with the best estimate of 0.637 kt TNT \cite{Stennett}.
Figure \ref{Fig:yield_comparison} shows a summary of all these independent estimates of the explosive yield.

\section{Conclusion}

Images can be powerful representations of events and in the case of explosions can contain rich information that can be studied.
The same methods for experimental analysis that students learn in their early years in physics laboratory experiences can serve to validate the modeling of a blast formation from a detonation.
Taylor's method for studying the blast from the first nuclear explosion at the Trinity test confirms that a complex relationship between the parameters describing the evolution of the explosion can be reduced to a simple linear fit leading to very accurate predictions.
The same method has been applied to the tragic ammonium nitrate explosion in Beirut.
The availability of plenty of footage of the explosion from different angles allows producing a new dataset that can be used for analyzing the evolution of the fireball as well as determining the yield of the explosion.
Despite the characteristics of this chemical explosion,
a remarkable agreement is found between the model and the data.
Fitting the model to the data allows determining an estimate for the energy yield of the Beirut explosion to be $2.3^{+1.1}_{-1.1}$ TJ or $0.6^{+0.3}_{-0.3}$ kt of TNT equivalent.
This estimate is in perfect agreement with the results from independent studies using different methods including infrasound, hydroacoustic, and seismic measurements as well as other audio-visual analysis of the blast.

\section*{Acknowledgments}
The author expresses his sympathies to all those affected by the events in Beirut on August 4, 2020 and thanks the people who recorded and made their audiovisual material available online.
The author is thankful to Newsflare Ltd. that provided the permission to include Figure \ref{Fig:frames}.
Comments and suggestions by the anonymous reviewers have significantly strengthened the manuscript and their observations are highly appreciated.
The author also thanks  C. Aouad, C. Cornell, O. Ram, and S. Rigby for their constructive feedback.
The whole analysis was conducted using Jupyter Notebooks \cite{jupyter} and all plots have been created using Matplotlib \cite{Matplotlib}.
This work was supported in part by the Indiana University Center for Spacetime Symmetries. 

\section*{Conflict of interest}
The author declares that he has no conflict of interest; 
referencing a social media profile or news reports does not indicate endorsement of the political or other views of that user, author, nor platform.


\begin{thebibliography}{99}
\bibitem{footnote}
This important observation was made by the two reviewers; the author was unaware of this enhancement of hemispherical changes leading to an overestimate of the yield in early versions of this manuscript.
\end{thebibliography}
\end{document}